\documentclass{amsart}

\usepackage{epsfig,amsfonts,amssymb,amsmath,amsthm,eucal,pinlabel}

\newtheorem{thm}{Theorem}[section]

\newtheorem{lemma}[thm]{Lemma}
\newtheorem{prop}[thm]{Proposition}
\newtheorem{cor}[thm]{Corollary}
\theoremstyle{remark}
\theoremstyle{remark}

\newenvironment{romanlist}
        {\begin{enumerate}
        }
        {\end{enumerate}}

\newcommand{\Z}{\mathbb Z}

\newcommand{\Hom}{\mathrm{Hom}}
\newcommand{\K}{\mathcal K}
\newcommand{\V}{\mathcal V}
\newcommand{\SI}{\Sigma}
\newcommand{\G}{\Gamma}
\newcommand{\GSS}{\Gamma\subset\Sigma}

\newcommand{\e}{\varepsilon}

\newcommand{\Pf}{\mbox{Pf}}
\newcommand{\Det}{\mbox{Det}}
\newcommand{\D}{\mathcal D}

\newcommand{\A}{\mbox{Arf}}

\begin{document}

\title{Dimers on surface graphs and spin structures. I}

\author{David Cimasoni}
\address{Department of Mathematics, UC Berkeley, 970 Evans Hall, Berkeley, CA 94720, USA}
\email{cimasoni@math.berkeley.edu}

\author{Nicolai Reshetikhin}
\email{reshetik@math.berkeley.edu}


\date{\today}

\begin{abstract}
Partition functions for dimers on closed oriented surfaces are known to be alternating sums of Pfaffians of
Kasteleyn matrices. In this paper, we obtain the formula for the coefficients in terms of discrete spin structures.
\end{abstract}

\maketitle

\tableofcontents

\section*{Introduction}

Dimer models on graphs have a long history in statistical mechanics
\cite{Kast1,McCoyW}. States in dimer models are
perfect matchings between vertices of the graph where only
adjacent vertices are matched. The probability of a state is
determined by assigning weights to edges.

Dimer models also have many interesting mathematical aspects
involving combinatorics, probability theory \cite{Kup,CKP},
real algebraic geometry \cite{KenOkS,KenOk}, etc... One
of the remarkable facts about dimer models is that the partition
function can be written as a linear combination of $2^{2g}$
Pfaffians of $N\times N$ matrices, where $N$ is the number of
vertices in the graph and $g$ the genus of a surface where the
graph can be embedded.

The matrices in the Pfaffian formula
for the dimer partition function are called Kasteleyn matrices.
They involve certain orientations of edges of the
graph known as Kasteleyn orientations. Two
Kasteleyn orientations are called equivalent if one can be obtained from
the other by a sequence of moves reversing orientations of all
edges adjacent to a vertex.

The number of non-equivalent Kasteleyn orientations of a surface
graph of genus $g$ is $2^{2g}$ and is equal to the number of
non-equivalent spin structures on the surface. The Pfaffian formula expresses the
partition function of the dimer model as an alternating sum of
Pfaffians of Kasteleyn operators, one for each equivalence class of
Kasteleyn orientations. This formula was proved in \cite{Kast1} for the
torus, and it was stated in \cite{Kast2} that for other surfaces, the partition
function of the dimer model is equal to the sum of $2^{2g}$
Pfaffians. The formal combinatorial proof of this fact and the
exact description of coefficients for all oriented surfaces first
appeared in \cite{L} and \cite{T} (see also \cite{Mis}). A combinatorial proof of such formula for
non-orientable surfaces can also be found in \cite{T}.

The partition function of free fermions on a Riemann surface of genus $g$ is also
a linear combination of $2^{2g}$ Pfaffians of Dirac operators.
Each term in this sum corresponds to a spin structure \cite{AGBMNV}. Assuming
that dimer models are discretizations of free fermions on Riemann
surfaces, one should expect a relation between Kasteleyn orientations
and spin structures and between the Kasteleyn operator for a given
Kasteleyn orientation and the Dirac operator in the corresponding spinor
bundle. Numerical evidence  relating the critical dimer model on a
square and triangular lattices in the thermodynamical limit with
Dirac operators can be found in \cite{CSMCoy} for $g=2$.

An explicit construction relating a spin structure on a surface
with a Kasteleyn orientation on a graph with dimer configuration was suggested in
\cite{Kup}. Furthermore, for bipartite graphs with critical weights, the
Kasteleyn operator can be naturally identified with a
discrete version of the Dirac operator \cite{Ken}. This gives
an interesting relation between dimer models and the theory of
discrete meromorphic functions \cite{M}.

In this paper we investigate further the relation between
Kasteleyn orientations and spin structures, and use this relation
to give a geometric proof of the Pfaffian formula for closed surfaces. 
Below is a
brief summary of our main results.

\medskip

Recall some basic notions. A dimer configuration on a graph $\G$ is a
perfect matching on vertices where matched vertices are connected
by edges. Given two such configurations $D$ and $D'$, set $\Delta(D,D')=(D\cup D')\setminus (D\cap D')$.
A surface graph is a graph $\Gamma$ embedded into a surface $\Sigma$ as
the 1-squeletton of a CW-decomposition of $\Sigma$. A
Kasteleyn orientation of a surface graph is an orientation of edges
of the graph, such that the product or relative orientations of
boundary edges of each face is negative (see Section \ref{section:Kasteleyn}).

One of our results is that any dimer configuration $D$ on a
surface graph $\GSS$ induces an isomorphism of affine
$H^1(\Sigma;\Z_2)$-spaces
\begin{equation}\label{corre}
\psi_D\colon(\K(\Gamma)/\sim)\longrightarrow
{\mathcal{Q}}(H_1(\SI;\Z_2),\cdot)\,,\quad[K]\longmapsto q_D^K
\end{equation}
from the set of equivalence classes of Kasteleyn orientations on
$\GSS$ onto the set of quadratic forms on $(H_1(\SI;\Z_2),\cdot)$,
where $\cdot$ denotes the intersection form on $\Sigma$.
Furthermore, $\psi_D=\psi_{D'}$ if and only if $D$ and $D'$ are
equivalent dimer configurations (that is, $\Delta(D,D')$ is zero in $H_1(\Sigma;\Z_2)$).

Since the affine space of spin structures on $\Sigma$ is canonically isomorphic to the
affine space of such quadratic forms, this establishes an isomorphism between equivalence
classes of Kasteleyn orientations and spin structures.

This correspondence implies easily the following identity.
Let $D_0$ be a fixed dimer configuration on a graph $\Gamma$. Realize $\Gamma$ as a surface graph $\GSS$ of genus $g$.
Let $K$ be any Kasteleyn orientation on $\GSS$, and let $A^K$ be the associated Kasteleyn matrix. Then,
\begin{equation}\label{Pfaff}
\Pf(A^K)=\e^K(D_0)\sum_{\alpha\in H_1(\Sigma;\Z_2)}(-1)^{q^K_{D_0}(\alpha)}\,Z_\alpha(D_0),
\end{equation}
where $q^K_{D_0}$ is the quadratic form associated to $K$ and $D_0$ via (\ref{corre}), $\e^K(D_0)$ is some sign depending on
$K$ and $D_0$, and
\[
Z_\alpha(D_0)=\sum_{D} w(D),
\]
the sum being on all dimer configurations $D$ such that $\Delta(D_0,D)$ is equal to $\alpha$ in  $H_1(\Sigma;\Z_2)$.

It follows that the partition function of a dimer model on $\Gamma$ is given by
\begin{equation}\label{partition}
Z=\frac{1}{2^{g}}\sum_{[K]}\A(q^{K}_{D_0})\e^K(D_0)\Pf(A^{K}),
\end{equation}
were the sum is taken over the $2^{2g}$ equivalence classes of Kasteleyn orientations on $\GSS$,
and $\A(q^{K}_{D_0})=\pm 1$ denotes the Arf invariant of the quadratic form $q^{K}_{D_0}$.
Note that the sign $\A(q^{K}_{D_0})\e^K(D_0)$ does not depend on $D_0$.

\medskip

The paper is organized as follows. In Section \ref{section:dimer}, we introduce the dimer model on a graph $\G$
and define composition cycles. Section \ref{section:surface} deals with dimers on surface graphs $\GSS$ and the definition
of an equivalence relation for dimer configurations on surface graphs. In Section \ref{section:Kasteleyn}, we recall the
definition of a Kasteleyn orientation. We then show that a surface graph $\GSS$ admits such an orientation if and only if
the number of vertices of $\Gamma$ is even. We prove that, in such a case, the set of equivalence classes of Kasteleyn orientations
on $\GSS$ is an affine $H^1(\Sigma;\Z_2)$-space. Finally, we give an algorithmic procedure for the construction of the $2^{2g}$
non-equivalent Kasteleyn orientations on a given surface graph $\GSS$ of genus $g$. The core of the paper lies
in Section \ref{section:spin}, where we establish the correspondence (\ref{corre}) stated above.
This result is used in Section \ref{section:Pfaff} to obtain equations (\ref{Pfaff}) and (\ref{partition}).
We also give a formula for the local correlation functions of a dimer model.
In the appendix, we collect formulae expressing dimer models in terms of Grassman integrals.

\subsection*{Acknowledgements}

We are grateful to Peter Teichner and Andrei Okounkov for
discussions, and to Richard Kenyon and Greg Kuperberg for useful
remarks and comments. We also thank Faye Yaeger who was kind to
type a large part of this paper. Finally, we thankfully acknowledge the hospitality
of the Department of Mathematics of the University of Aarhus, via the Niels Bohr initiative.
The work of D.C.\ was 
supported by the Swiss National Science Foundation. The work of N.R.\ was supported
by the NSF grant DMS-0307599, by the CRDF grant RUM1-2622, and by
the Humboldt foundation.

\section{The dimer model}\label{section:dimer}

\subsection{Dimer configurations and composition cycles on graphs}

\begin{figure}[htbp]
 \begin{center}
    \epsfig{figure=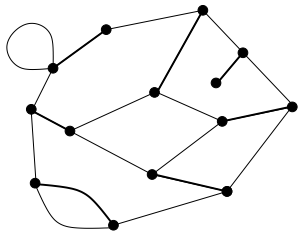,height=3cm}
    \caption{A dimer configuration on a graph.}
    \label{ex-dimer}
  \end{center}
\end{figure}

Let $\Gamma$ be a finite connected graph. A {\em perfect matching on $\G$\/} is a choice of edges of $\G$ such that each vertex of $\G$ is
adjacent to exactly one of these edges. In statistical mechanics, a perfect matching on $\G$ is also known as a {\em dimer configuration on $\G$\/}.
The edges of the perfect matching are called {\em dimers\/}. An example of a dimer configuration on a graph is given in Figure \ref{ex-dimer}.

In order to have a perfect matching, a graph $\G$ clearly needs to
have an even number of vertices. However, there are connected
graphs with an even number of vertices but no perfect matching. We
refer to \cite{LP} for combinatorial aspects of matchings.
Throughout the paper and unless otherwise stated, we will only
consider finite graphs which admit perfect matchings. In
particular, all the graphs will have an even number of vertices.

\begin{figure}[htbp]
 \begin{center}
    \epsfig{figure=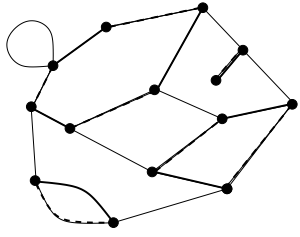,height=3cm}
    \caption{An example of composition cycles: dimers from $D$ are in solid, and dimers from $D'$ are in traced lines. On this example, there is one
$(D,D')$-composition cycle of length 2, one of length 4, and one of length 6.}
    \label{comp-cycles}
  \end{center}
\end{figure}

Given two dimer configurations $D$ and $D'$ on a graph $\G$, consider the subgraph of $\G$ given by the symmetric difference
$(D\cup D')\backslash(D\cap D')$.
The connected components of this subgraph are called {\em $(D,D')$-composition cycles\/} or simply {\em composition cycles\/}.
Clearly, each composition cycle is a simple closed curve of even length. This is illustrated in Figure \ref{comp-cycles},
where the two dimer configurations are shown in black and traced lines.

\subsection{Edge weight system}

Let $\D(\G)$ denote the set of dimer configurations on a graph $\G$. A {\em weight system\/} on $\D(\G)$ is a positive real-valued
function on this set. A weight system $w$ defines a probability distribution on all dimer configurations:
\[
\mbox{Prob}(D)=\frac{w(D)}{Z(\G; w)},
\]
where
\[
Z(\G; w)=\sum_D w(D)
\]
is the partition function. This probabilistic measure is the {\em Gibbs measure\/} for the dimer model on the graph $\G$ with the weight system $w$.

We shall focus on a particular type of weight system called {\em edge weight system}. Assign to each edge $e$ of $\G$ a positive real number $w(e)$,
called the {\em weight\/} of the edge $e$. The associated edge weight system on $\D(\G)$ is given by
\[
w(D)=\prod_{e\in D}w(e),
\]
where the product is over all edges occupied by dimers of $D$. In statistical mechanics, these weights are called {\em Boltzmann weights\/}.
Their physical meaning is
\[
w(e)=\exp\Big(-\frac{E(e)}{T}\Big),
\]
where $E(e)$ is the energy of dimer occupying the edge $e$ and $T$ is the temperature.

\subsection{Local correlation functions}

Let $e$ be an edge of $\G$. The characteristic function of $e$ is the function $\sigma_e$ on $\D(\G)$ given by
\[
\sigma_e(D)=
\begin{cases}
1 & \text{if $e\in D$;} \\
0 & \text{otherwise.}
\end{cases}
\]

The expectation values of products of characteristic functions are called {\em local correlation functions\/}, or {\em dimer-dimer correlation functions\/}:
\[
<\sigma_{e_1}\cdots \sigma_{e_k}>=\frac{Z(e_1,\dots, e_k;\G;w)}{Z(\G;w)},
\]
where
\[
Z(e_1,\dots, e_k;\G;w)=\sum_D \prod_{e\in D}w(e)\prod_{i=1}^k\sigma_{e_i}(D)=\sum_{D\ni e_1,\dots,e_k}w(D).
\]
Note that $<\sigma_{e_1}\cdots\sigma_{e_k}>=0$ if some edges $e_i\neq e_j$ share a common vertex. Note also that $\sigma_e\cdot\sigma_e=\sigma_e$.
Therefore, it may be assumed that $e_i\neq e_j$ for $i\neq j$.

For any dimer configuration $D$,
\[
\big<\prod_{e\in D}\sigma_e\big>=\frac{w(D)}{Z(\G;w)},
\]
so we can reconstruct the weight system if we know all local
correlation functions. In this sense, local correlation functions
carry all the information about the Gibbs measure.

\section{Dimers on surface graphs}\label{section:surface}

\subsection{Surface graphs}

Let $\Sigma$ be a connected oriented closed surface. By a {\em surface graph\/}, we mean a graph $\Gamma$ embedded in $\Sigma$ as the
1-squeleton of a cellular decomposition $X$ of $\Sigma$. We shall assume throughout the paper that the surface $\Sigma$ is endowed with the
counter-clockwise orientation.

Any finite connected graph can be realized as a surface graph. Indeed, such a graph $\G$ always embeds in a closed oriented surface of genus $g$, for
$g$ sufficiently large. If the genus is minimal, one easily checks that $\G$ induces a cellular decomposition of the surface $\SI$.

We will be interested mostly in large graphs. In this paper we
will focus on graphs embedded into a surface of fixed genus.

\subsection{Equivalent dimer configurations}

A dimer configuration on a surface graph $\GSS$ is simply a dimer configuration on the graph $\G$. Such a dimer configuration can be regarded as a
1-chain in the cellular chain complex of $X$ with $\Z_2$-coefficients:
\[
c_D=\sum_{e\in D}e\in C_1(X;\Z_2).
\]
By definition, $\partial c_D=\sum_{v\in\G}v\in C_0(X;\Z_2)$, the sum being on all vertices $v$ of $\Gamma$. Therefore, given any pair of dimer
configurations $D$ and $D'$ on $\GSS$, $c_D+c_{D'}$ is a 1-cycle:
\[
\partial(c_D+c_{D'})=\partial c_D+\partial c_{D'}=\sum_{v\in\G}(v+v)=0\in C_0(X;\Z_2).
\]
This 1-cycle is nothing but the union of all $(D,D')$-composition cycles. Let $\Delta(D,D')$ denote its homology class in
$H_1(X;\Z_2)=H_1(\SI;\Z_2)$. We shall say that two dimer configurations $D$ and $D'$ are {\em equivalent\/} if $\Delta(D,D')=0$ in $H_1(\SI;\Z_2)$.

Note that these concepts make perfect sense when $\G$ is the 1-squeleton of any CW-complex, not necessarily the cellular
decomposition of an oriented closed surface.

\section{Kasteleyn orientations on surface graphs}\label{section:Kasteleyn}

\begin{figure}[htbp]
 \begin{center}
    \epsfig{figure=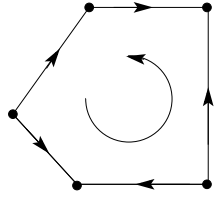,height=2.5cm}
        \caption{A Kasteleyn orientation on the boundary edges of a face.}
        \label{Kast}
    \end{center}
\end{figure}

Let $\GSS$ be a surface graph. The counter-clockwise orientation of $\SI$ induces an orientation on each 2-cell, or {\em face\/} of $X$.
An orientation $K$ of the edges of $\G$ is called a {\em Kasteleyn orientation\/} if for each face $f$ of $X$,
\begin{equation}\label{K-prop}
\prod_{e\in \partial f}\e^K_f(e)=-1,
\end{equation}
where the product is taken over all boundary edges of $f$, and
\[
\e^K_f(e)=\begin{cases}
\phantom{-}1 & \text{if $e$ is oriented by $K$ as the oriented boundary of the face $f$;} \\
-1 & \text{otherwise.}
\end{cases}
\]
This is illustrated in Figure \ref{Kast}.

\begin{figure}[htbp]
 \begin{center}
    \epsfig{figure=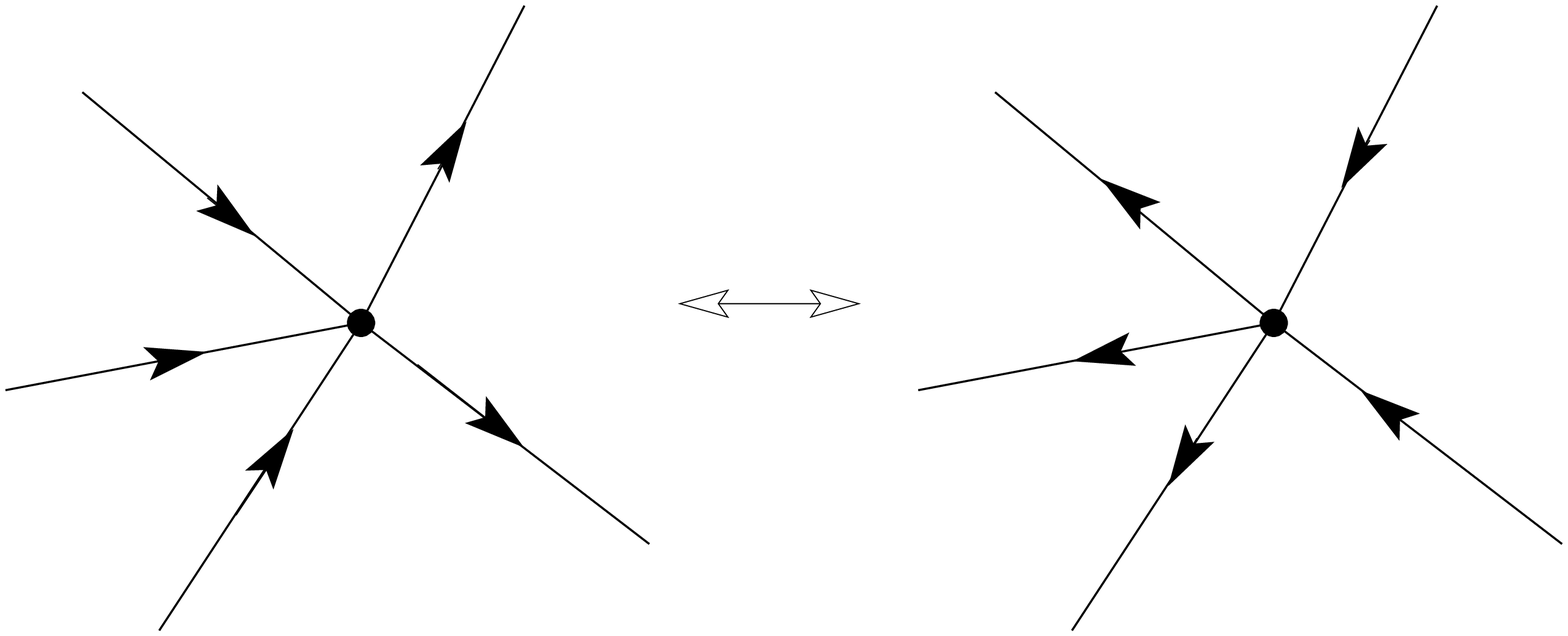,height=3cm}
    \caption{Orientation change at a vertex.}
    \label{flip-vert}
  \end{center}
\end{figure}

Define the operation of {\em orientation changing at a vertex\/} as the one which flips the orientation of all the edges adjacent to this
vertex, as illustrated in Figure \ref{flip-vert}. It is clear that such an operation brings a Kasteleyn orientation
to a Kasteleyn orientation. Let us say that two Kasteleyn orientations are {\em equivalent\/} if they are
obtained one from the other by a sequence of orientation changes at vertices.

\subsection{Existence of a Kasteleyn orientation}

\begin{thm}\label{thm:1}
There exists a Kasteleyn orientation on a surface graph $\Gamma\subset\Sigma$ if and only if the number of vertices of $\Gamma$ is even.
\end{thm}
\begin{proof}
Let $\omega$ be any orientation of the edges of $\Gamma$, and let $c_{\omega}\in C^2(X;\Z_2)$ be defined by the equation
$$
(-1)^{c_\omega(f)}=-\prod_{e\in\partial f}\e^\omega_f(e).
$$
Note that $\omega$ is Kasteleyn if and only if $c_\omega=0$.
Let $V$, $E$ and $F$ denote the number of vertices,
edges and faces in $X$, respectively. Since $\Sigma$ is closed, its Euler characteristic is even, and we get the equality mod~$2$
\[
0 = \chi(\Sigma) = V+E+F = V+\sum_{f\in F} c_{\omega}(f).
\]
Here, each edge $e$ contributes to the number $c_{\omega}(f)$ where $f$ is the face whose oriented boundary contains $e$ with the orientation
opposite to $\omega$. Therefore, the number of faces $f$ such that $c_{\omega}(f) = 1$ has the parity of $V$. Hence, if $V$ is even, then
$c_{\omega}(f) = 1$ for an even number of faces, so $c_{\omega}$ is a $2$-coboundary. In other words, $c_{\omega} = \delta\sigma$ for some
$\sigma\in C^1(X;\Z_2)$. Let $K$ be the orientation of the edges of $\Gamma$ which agrees with $\omega$ on $e$ if and only if $\sigma(e)=0$.
Clearly, $c_K = 0$, so $K$ is a Kasteleyn orientation. Conversely, let us assume that there is a Kasteleyn orientation $K$.
This means that $c_K(f) = 1$ for none of the faces. By the argument above, $V$ is even.
\end{proof}

A more constructive proof of this result will be given in Section \ref{subsection:algorithm}.

\subsection{Uniqueness of Kasteleyn orientations}

Let $V$ be a vector space. Recall that an {\em affine $V$-space\/} is a set $S$ endowed with a map
$S\times S\to V $, $(a,b)\mapsto a-b$ such that:
\begin{romanlist}
\item{for every $a$, $b$ and $c$ in $S$, we have $(a-b)+(b-c)=a-c$~;}
\item{for every $b$ in $S$, the map $S \to V$ given by $a \mapsto a - b$ is a bijection.}
\end{romanlist}
In other words, an affine $V$-space is a {\em $V$-torsor\/}: it is a set endowed with a freely transitive action of the abelian group $V$.

\begin{thm}\label{thm:aff}
Let $\Gamma\subset\Sigma$ be a surface graph with an even number of vertices.
Then, the set of equivalence classes of Kasteleyn orientations on $\Gamma\subset\Sigma$ is an affine $H^1(\Sigma;\Z_2)$-space.
\end{thm}
\begin{cor}\label{2^2g}
There are exactly $2^{2g}$ equivalence classes of Kasteleyn orientations on $\Gamma\subset\Sigma$, where $g$ denotes the genus of $\Sigma$.\qed
\end{cor}
\begin{proof}[Proof of the theorem.]
Let $\K(\Gamma)$ denote the set of Kasteleyn orientations on $\Gamma \subset \Sigma$. By Theorem \ref{thm:1}, it is non-empty.
Consider the map $\vartheta\colon \K(\Gamma)\times \K(\Gamma)\to C^1(X;\Z_2)$ given by $\vartheta_{K,K'}(e) = 0$ if
$K$ and $K'$ agree on the edge $e$, and $\vartheta_{K,K'}(e) = 1$ otherwise. Since $K$ and $K'$ are Kasteleyn orientations,
\[
(-1)^{\vartheta_{K,K'}(\partial f)}=\prod_{e\in\partial f}(-1)^{\vartheta_{K,K'}(e)}=
\prod_{e\in\partial f}\e^K_f(e)\cdot\prod_{e\in\partial f}\e^{K'}_f(e)=(-1)(-1)= 1
\]
for any face $f$. Therefore, $\delta\vartheta_{K,K'}(f) = \vartheta_{K,K'}(\partial f) = 0$, that is, $\vartheta_{K,K'}$ is a $1$-cocycle. Thus, we
get a map $\K(\Gamma)\times\K(\Gamma)\stackrel{\vartheta}{\to} H^1(\Sigma;\Z_2)$.  Note that
$\vartheta_{K,K'}+\vartheta_{K',K''}= \vartheta_{K,K''}$ for any $K,K',K'' \in \K(\Gamma)$.  Also, one easily checks that $\vartheta_{K,K'}=0$ in $H^1(\Sigma;\Z_2)$ if and only if there is a sequence of vertices such that $K$ is obtained from $K'$ by reversing the orientation around all these vertices, i.e., if and only if $K \sim K'$.  It follows that we have a map
\[
(\K(\Gamma)/\sim)\times(\K(\Gamma)/\sim)\longrightarrow H^1(\Sigma;\Z_2),\quad([K],[K'])\mapsto [K]-[K']:=[\vartheta_{K,K'}]
\]
such that for any $K'$ in $\K(\Gamma)$, the map $(\K(\Gamma)/\sim)\to H^1(\Sigma;\Z_2)$ given by $[K]\mapsto[K]-[K']$ is injective. Finally,
let us check that this map is onto.
Fix a class in $H^1(\Sigma;\Z_2)$, and represent it by some 1-cycle $\sigma\in Z^1(X;\Z_2)$.
Let $K$ be the same orientation as $K'$ whenever $\sigma(e) = 0$, and the opposite when $\sigma(e) = 1$.  Obviously,
$\vartheta_{K,K'}=\sigma$. Furthermore, $K$ is Kasteleyn since $K'$ is and $\delta\sigma = 0$. Indeed, given a face $f$,
\[
0 = (\delta\sigma)(f) = \sigma(\partial f) = \vartheta_{K,K'}(\partial f),
\]
so $1=\prod_{e\in\partial f}(-1)^{\vartheta_{K,K'}(e)}=(-1)\prod_{e\in\partial f}\e^K_f(e)$.  This concludes the proof.
\end{proof}

\subsection{How to construct Kasteleyn orientations}\label{subsection:algorithm}

\begin{figure}[htbp]
\labellist\small\hair 2.5pt
\pinlabel {0.} at 80 930
\pinlabel {1.} at 850 930
\pinlabel {2.} at -10 480
\pinlabel {3.} at 600 480
\pinlabel {4.} at 1210 480
\pinlabel {$\GSS$} at 620 610
\pinlabel {$\alpha_1$} at 1323 780
\pinlabel {$\alpha_2$} at 1205 930
\pinlabel {$T$} at 85 20
\pinlabel {$\G'\subset S^2$} at 410 35
\pinlabel {$e_*$} at 685 33
\endlabellist
\centerline{\psfig{file=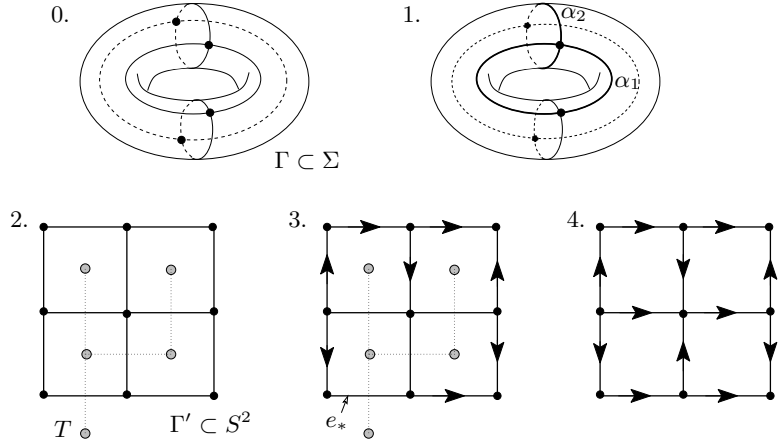,height=6cm}}
\caption{An example of the explicit construction of a Kasteleyn orientation.}
\label{fig:algorithm}
\end{figure}

Given a surface graph $\GSS$ with an even number of vertices, we know that there are exactly $2^{2g}$ non-equivalent Kasteleyn orientations on $\G$.
However, the proof given above is not really constructive. For this reason, we now give an algorithm for the construction of these Kasteleyn orientations.
The successive steps of the algorithm are illustrated in Figure \ref{fig:algorithm}.
\begin{enumerate}\setcounter{enumi}{-1}
\item Let $\GSS$ be a surface graph with an even number of vertices, and let $g$ denote the genus of $\Sigma$.
\item Consider a system $\alpha=\alpha_1\cup\dots\cup\alpha_{2g}$ of simple closed curves on $\Gamma$ such that $\SI$ cut along $\alpha$ is a 2-disc $\SI'$.
(Note that such curves exist since $\G$ induces a cellular decomposition of $\SI$.)
\item The surface graph $\GSS$ induces a graph $\G'\subset\SI'$ which is the 1-squeleton of a cellular decomposition of the 2-sphere $S^2$.
Fix a spanning tree $T$ of the graph dual to the surface graph $\G'\subset S^2$, rooted at the vertex corresponding to the face $S^2\backslash\SI'$.
\item Orient the edges of $\G'$ which do not intersect $T$ respecting the following condition: whenever two edges of $\G'$ in $\partial\SI'$
are identified in $\SI$, their orientations agree. Now, every edge in $\partial\SI'$ is oriented, except one. Let us denote it by $e_*$.
\item Orient edges of $\G'$ such that the Kasteleyn condition (\ref{K-prop}) holds for the faces corresponding to the leaves of $T$.
Moving down the tree (from the leaves to the root), orient each crossing edge such that the Kasteleyn condition holds for all faces left behind.
\end{enumerate}
This gives a Kasteleyn orientation of $\G'\subset\SI'$. By the condition in step 3, it induces a Kasteleyn orientation on $\GSS$, provided that
the orientation of the last edge $e_*$ satisfies this condition. It turns out to be the case if and only if the number of vertices of $\G$ is even.
(This is an easy consequence of Theorem \ref{thm:1}.) Therefore, we have constructed a Kasteleyn orientation $K$ on $\GSS$. To obtain the $2^{2g}$
non-equivalent ones, proceed as follows.
\begin{enumerate}\setcounter{enumi}{4}
\item Consider a family of simple closed curves $\beta_1,\dots,\beta_{2g}$ on $\SI$ avoiding the vertices of $\G$, and forming a basis of
$H_1(\Sigma;\Z_2)$.
\item Consider the Kasteleyn orientation $K$, and some subset $I\subset\{1,\dots,2g\}$. For all $i\in I$, change the orientation
of all the edges in $\G$ that intersect $\beta_i$.
\end{enumerate}
The resulting orientation $K_I$ is clearly Kasteleyn, as $\partial f\cdot\beta_i$ is even for every face $f$ and index $i$. Furthermore, one easily
checks that $K_I$ and $K_J$ are non-equivalent if $I\neq J$. Hence, we have constructed the $2^{2g}$ non-equivalent Kasteleyn orientations on $\GSS$.

\section{Kasteleyn orientations as discrete spin structures}\label{section:spin}

We saw in Corollary \ref{2^2g} that there are exactly $2^{2g}$ non-equivalent Kasteleyn orientations on a surface graph $\GSS$, where $g$ denotes the
genus of $\SI$. It is known that this is also the number of non-equivalent spin structures on $\Sigma$. This relation between the number of
Kasteleyn orientations and the number of spin structures is not accidental. Kasteleyn orientations of surface graphs can be
regarded as discrete versions of spin structures. This statement will be made precise in the present section. (See in particular Corollary \ref{cor:spin}.)

\subsection{The quadratic form associated to a Kasteleyn orientation}

Let $V$ be a finite dimensional vector space over the field $\Z_2$, and let $\varphi\colon V\times V\to \Z_2$ be a fixed bilinear form.
Recall that a function $q\colon V\to \Z_2$ is a {\em quadratic form\/} on $(V,\varphi)$ if
\[
q(x+y)=q(x)+q(y)+\varphi(x,y)
\]
for all $x,y\in V$. Note that the difference (that is, the sum) of two quadratic forms on $(V,\varphi)$ is a linear form on $V$.
Therefore, one easily checks that the set $\mathcal{Q}(V,\varphi)$ of quadratic forms on $(V,\varphi)$ is an affine $V^*$-space,
where $V^*$ denotes the dual of $V$.

Fix a Kasteleyn orientation $K$ on a surface graph $\Gamma\subset\Sigma$. Given an oriented simple closed curve $C$ on $\Gamma$, set
$$
\e^K(C)=\prod_{e\in C}\e^K_C(e),
$$
where $\e^K_C(e)$ is equal to $+1$ (resp. $-1$) if the orientations on the edge $e$ given by $C$ and $K$ agree (resp. do not agree).
For a fixed dimer configuration $D$ on $\Gamma$, let $\ell_D(C)$ denote the number of vertices $v$ in $C$ whose adjacent dimer of $D$ sticks out
to the left of $C$ in $\Sigma$.

\begin{thm}\label{thm:spin}
Given a class $\alpha\in H_1(\SI;\Z_2)$, represent it by oriented simple closed curves $C_1,\dots,C_m$ in $\G$. If $K$ is a Kasteleyn orientation on
$\GSS$, then the function $q^K_D\colon H_1(\Sigma;\Z_2)\to\Z_2$ given by
\[
(-1)^{q^K_D(\alpha)}=(-1)^{\sum_{i<j}C_i\cdot C_j}\prod_{i=1}^m(-\e^K(C_i))(-1)^{\ell_D(C_i)}
\]
is a well-defined quadratic form on $(H_1(\SI;\Z_2),\cdot)$, where $\cdot$ denotes the intersection form.
\end{thm}

We postpone the proof of this result to the next subsections. Let us first investigate some of its consequences.

\begin{prop}\label{prop:comp}
$(i)$ Let $D$ be a fixed dimer configuration on $\Gamma$. If $K$ and $K'$ are two Kasteleyn orientations on $\Gamma\subset\Sigma$, then
$q^K_D-q^{K'}_D$ maps to $[K]-[K']$ via the canonical isomorphism $\Hom(H_1(\Sigma;\Z_2);\Z_2)=H^1(\Sigma;\Z_2)$.

\noindent $(ii)$  Let $K$ be a fixed Kasteleyn orientations on $\Gamma\subset\Sigma$. If $D$ and $D'$ are two dimer configurations
on $\Gamma$, then $q^K_D-q^K_{D'}\in \Hom(H_1(\Sigma;\Z_2);\Z_2)$ is given by $\alpha\mapsto\alpha\cdot\Delta(D,D')$.
\end{prop}
\begin{proof}
Let $C$ be a simple closed curve in $\Gamma$ representing a class $\alpha$ in $H_1(\Sigma;\Z_2)$. By definition,
\[
(-1)^{(q^K_D-q^{K'}_D)(\alpha)}=\prod_{e\in C}\e^K_C(e)\prod_{e\in C}\e^{K'}_C(e)=\prod_{e\in C}(-1)^{\vartheta_{K,K'}(e)}=(-1)^{\vartheta_{K,K'}(C)}.
\]
This proves the first point. To check the second one, observe that
\[
(-1)^{(q^K_D-q^K_{D'})(\alpha)}=(-1)^{\ell_D(C)+\ell_{D'}(C)}.
\]
Clearly, $\ell_D(C)+\ell_{D'}(C)\equiv C\cdot\Delta(D,D')\pmod{2}$, giving the proposition.
\end{proof}

\begin{cor}\label{cor:quadratic}
Any dimer configuration $D$ on a surface graph $\GSS$ induces an isomorphism of affine $H^1(\Sigma;\Z_2)$-spaces
\[
\psi_D\colon(\K(\Gamma)/\sim)\longrightarrow {\mathcal{Q}}(H_1(\SI;\Z_2),\cdot)\,,\quad[K]\longmapsto q_D^K
\]
from the set of equivalence classes of Kasteleyn orientations on $\GSS$ onto the set of quadratic forms on $(H_1(\SI;\Z_2),\cdot)$.
Furthermore, $\psi_D=\psi_{D'}$ if and only if $D$ and $D'$ are equivalent dimer configurations.
\end{cor}
\begin{proof}
The first part of Proposition \ref{prop:comp} exactly states that $\psi_D$ is an isomorphism of affine $H^1(\Sigma;\Z_2)$-spaces. By the second part,
$\psi_D=\psi_{D'}$ if and only if the homomorphism $H_1(\SI;\Z_2)\to\Z_2$ given by the intersection with $\Delta(D,D')$ is zero. By Poincar\'e duality,
this is the case if and only if $\Delta(D,D')=0$.
\end{proof}

\subsection{Spin structures on surfaces}

Recall that the fundamental group of $SO(n)$ is infinite cyclic if $n=2$ and cyclic of order 2 if $n\ge 3$. Hence, $SO(n)$ admits a canonical 2-fold
cover, denoted by $Spin(n)\to SO(n)$.

Let $M$ be an oriented $n$-dimensional Riemannian manifold, and let $P_{SO}\to M$ be the principal
$SO(n)$-bundle associated to its tangent bundle. A {\em spin structure on $M$\/} is a principal $Spin(n)$-bundle $P\to M$ together with a
2-fold covering map $P\to P_{SO}$ which restricts to the covering map $Spin(n)\to SO(n)$ on each fiber.
Equivalently, a spin structure on $M$ is a cohomology class
$\xi\in H^1(P_{SO};\Z_2)$ whose restriction to each fiber $F$
gives the generator of the cyclic group $H^1(F;\Z_2)$.

It is well-known that such a spin structure exists if and only
if the second Stiefel-Whitney class of $M$ vanishes. In such a case, the set $\mathcal{S}(M)$ of spin structures on $M$ is endowed with a natural structure
of affine $H^1(M;\Z_2)$-space.

\medskip

The 2-dimensional case is particularly easy to deal with for several reasons. First of all, any compact orientable surface $\SI$ admits a spin structure,
as its second Stiefel-Whitney class is always zero. Furthermore, spin structures can be constructed using a certain class a vector fields, that we
now describe.

Let $f$ be a non-vanishing vector field on $\Sigma\setminus\sigma$, where $\sigma$ is some finite subset of $\SI$. Recall that the
{\em index of the singularity $x\in\sigma$\/} of $f$ is defined as the degree of the circle map $t\mapsto f(\gamma_x(t))/|f(\gamma_x(t))|$,
where $\gamma_x\colon S^1\to\SI\setminus\sigma$ is a (counter-clockwise) parametrization of a simple closed curve separating $x$ from the other singularities of $f$. Let us denote by $\V_{ev}(\SI)$ the set
of vector fields on $\SI$ with only even index singularities. We claim that any such vector field $f$ defines a spin structure $\xi_f$ on $\SI$.
Indeed, consider a 1-cycle $c$ in $P_{SO}$ (that is, a closed framed curve in $\Sigma$) and let us assume that $c$ avoids the singularities of $f$.
Then, let $\xi_f(c)\in\Z_2$ be the winding number modulo 2 of $f$ along $c$ with respect to the framing of $c$. Since all the singularities of $f$ have
even  index, $\xi_f(c)=0$ if $c$ is a 1-boundary, and $\xi_f(c)=1$ if $c$ is a small simple closed curve with tangential framing. Therefore, it induces
a well-defined cohomology class $\xi_f\in \Hom(H_1(P_{SO};\Z_2);\Z_2)=H^1(P_{SO};\Z_2)$ which restricts to the generator of the cohomology of the fibers.
So $\xi_f$ is a spin structure on $\SI$, and we have a map
\[
\V_{ev}(\SI)\longrightarrow{\mathcal S}(\SI)\,,\quad f\longmapsto \xi_f.
\]

We shall need one last result about spin structures on surfaces, due to D. Johnson \cite{Joh}. Given a spin structure $\xi\in{\mathcal S}(\SI)$, let
$q_\xi\colon H_1(\SI;\Z_2)\to\Z_2$ be the function defined as follows. Represent $\alpha\in H_1(\SI;\Z_2)$ by a collection of disjoint regular simple
closed curves $\gamma_1,\dots,\gamma_m\colon S^1\hookrightarrow\SI$. For all $i$ and all $t\in S^1$, complete the unit tangent vector
$\dot{\gamma_i}(t)/|\dot{\gamma_i}(t)|$ to a positive orthonormal basis of $T_{\gamma_i(t)}\SI$. This gives disjoint framed closed curves in $\SI$,
that is, a 1-cycle $c$ in $P_{SO}$. Set $q_\xi(\alpha)=\xi(c)+m$. Johnson's theorem asserts that $q_\xi$ is a well-defined quadratic form on
$(H_1(\Sigma;\Z_2),\cdot)$, where $\cdot$ denotes the intersection form. Furthermore, the map
\[
{\mathcal S}(\SI)\longrightarrow{\mathcal Q}(H_1(\Sigma;\Z_2),\cdot)\,,\quad \xi\longmapsto q_\xi
\]
is an isomorphism of affine $H^1(\SI;\Z_2)$-spaces.

\subsection{Proof of Theorem \ref{thm:spin}}

\begin{figure}[htbp]
 \begin{center}
    \epsfig{figure=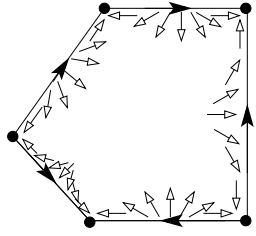,height=3cm}
        \caption{Kuperberg's construction.}
        \label{fig:vector}
    \end{center}
\end{figure}

Let $\GSS$ be a surface graph, with $\Sigma$ right-hand oriented. Given a Kasteleyn orientation $K$ and a dimer configuration $D$
on $\GSS$, Kuperberg \cite{Kup} constructs a vector field $f(K,D)\in\V_{ev}(\SI)$ as follows. Around each vertex of $\Gamma$, make the vectors point to
the vertex. At the middle of each edge, make the vector point 90 degrees clockwise relative to the orientation $K$ of the edge. Extend this continuously
to the whole edges, as described in Figure \ref{fig:vector}.

The Kasteleyn condition (\ref{K-prop}) ensures that the vector field extends to the faces with one singularity of even index in the interior of each face.
However, this vector field $\tilde{f}(K)$ has an odd index singularity at each vertex of $\G$. This is where the dimer configuration $D$ enters the game:
contract the odd index singularities in pairs along the dimers of $D$. The resulting vector field $f(K,D)$ has even index singularities: one in the interior
of each face of $\Sigma$, and one in the middle of each dimer of $D$.

\begin{figure}[htbp]
\labellist\small\hair 2.5pt
\pinlabel {$\gamma$} at 340 500
\pinlabel {$C$} at 440 485
\pinlabel {$\gamma_0$} at 850 500
\pinlabel {$C$} at 955 485
\pinlabel {$\tilde\gamma$} at 1408 500
\pinlabel {$C$} at 1510 485
\endlabellist
\centerline{\psfig{file=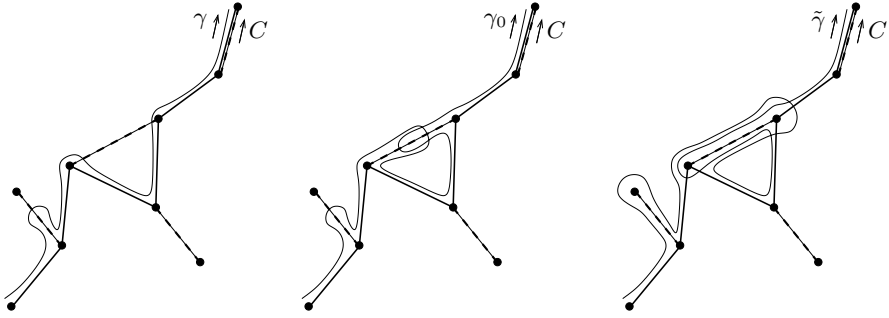,height=4.1cm}}
\caption{The oriented simple closed curve $C\subset\G$ and its associated regular curves $\gamma$, $\gamma_0$ and $\tilde\gamma$.
The curve $C$ is in solid, the dimers are in traced lines.}
\label{fig:proofthm}
\end{figure}

Gathering the results of the previous section and Kuperberg's construction, we get the following composition of maps:
\[
\K(\G)\times\D(\G)\longrightarrow\V_{ev}(\SI)\longrightarrow{\mathcal S}(\SI)\longrightarrow{\mathcal Q}(H_1(\Sigma;\Z_2),\cdot)\,,
\quad (K,D)\longmapsto q_{\xi_{f(K,D)}}.
\]
We are left with the proof that, given any Kasteleyn orientation $K$ and dimer configuration $D$, the resulting quadratic form $q=q_{\xi_{f(K,D)}}$
coincides with the function $q_D^K$ defined in the statement of Theorem \ref{thm:spin}.
So, given $\alpha\in H_1(\SI;\Z_2)$, represent it by a collection of simple closed curves $C_1,\dots,C_m$ in $\G$.
(This is always possible as $\G$ induces a cellular decomposition of $\SI$.)
Since $q$ is a quadratic form and $\alpha=\sum_{i=1}^m[C_i]$,
\[
(-1)^{q(\alpha)}=(-1)^{\sum_{i<j}C_i\cdot C_j}\prod_{i=1}^m(-1)^{q([C_i])}.
\]
Therefore, we just need to check that if $C$ is an oriented simple closed curve in $\G$, then $(-1)^{q([C])+1}=\e^K(C)(-1)^{\ell_D(C)}$.
Consider the oriented regular curve $\gamma$ in $\SI$ which follows $C$ slightly on its left, and goes around the middle of each dimer it meets,
except if it meets the same dimer twice. In this case, $\gamma$ stays close to $C$, as illustrated in Figure \ref{fig:proofthm}.
Clearly, $\gamma$ is a regular oriented simple closed curve in $\SI$. Furthermore, it is homologous to $C$ and it avoids
all the singularities of $f(K,D)$. We now have to check that the winding number $\omega$ of $f(K,D)$ along $\gamma$ with respect to its tangential
framing satisfies $(-1)^\omega=\e^K(C)(-1)^{\ell_D(C)}$. One easily check that $\omega$ is equal $\pmod{2}$ to the winding number $\omega_0$ of
$f(K,D)$ along $\gamma_0$, where $\gamma_0$ is the regular curve which goes around the middle of each dimer it meets, including the ones it meets twice.
By construction of $f(K,D)$, $\omega_0$ is equal to the winding number $\tilde\omega$ of $\tilde f(K)$ along $\tilde\gamma$, where $\tilde\gamma$ is the regular curve which avoids all the dimers of $D$ and all the vertices of $\G$, as described in Figure \ref{fig:proofthm}.

The latter winding number $\tilde\omega$ can be computed locally by cutting $\tilde\gamma$ into pieces: one piece $\tilde\gamma_e$ for each edge $e$ of
$C$, and one piece $\tilde\gamma_v$ for each vertex $v$ of $\G$ adjacent to a dimer of $D$ sticking out to the left of $C$.
The theorem now follows from the following case study.

\begin{figure}[htbp]
\labellist\small\hair 2.5pt
\pinlabel {1.} at 0 350
\pinlabel {2.} at 400 350
\pinlabel {3.} at 790 350
\pinlabel {$C$} at 225 275
\pinlabel {$C$} at 615 275
\pinlabel {$C$} at 883 20
\pinlabel {$e$} at 110 120
\pinlabel {$e$} at 500 120
\pinlabel {$v$} at 870 60
\endlabellist
\centerline{\psfig{file=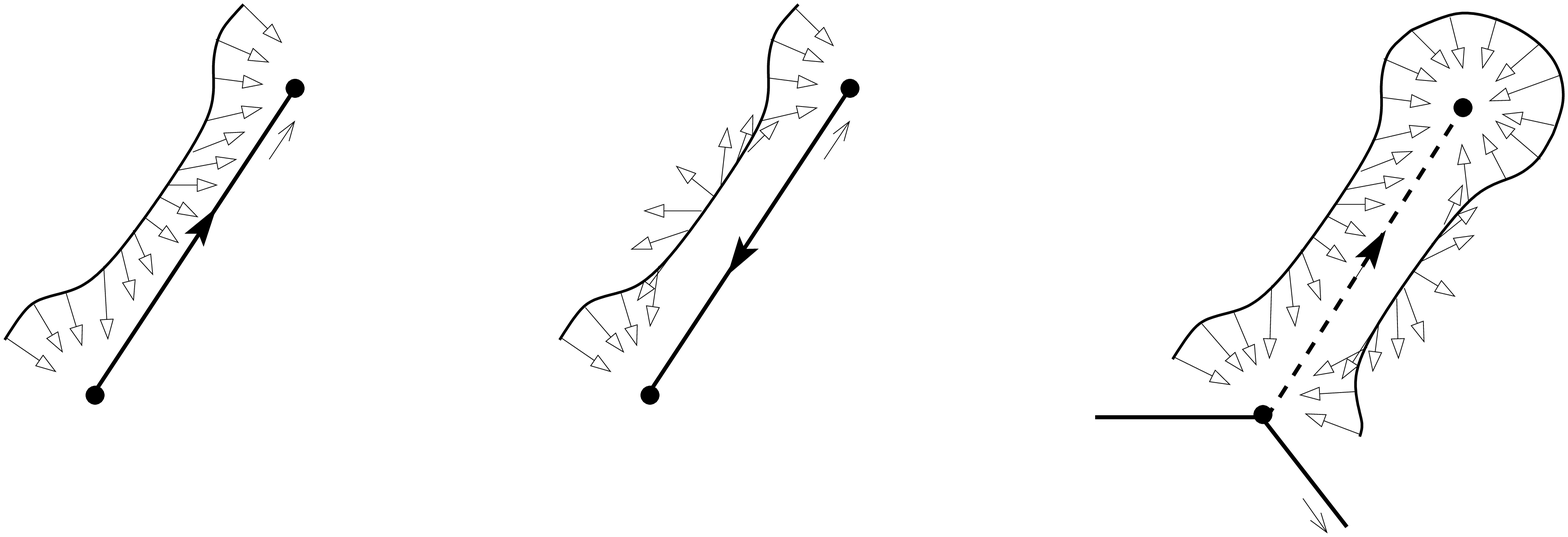,height=4cm}}
\caption{Computation of the local winding numbers.}
\label{3cases}
\end{figure}

1. Let us assume that $e$ is an edge of $C$ such that $\e^K_C(e)=+1$. In this case, the vector field $\tilde f(K)$ along $\tilde\gamma_e$ in the
tangential framing of $\tilde\gamma_e$ defines a curve which is homotopically trivial, as illustrated in Figure \ref{3cases}.
Hence, its contribution to $\tilde\omega$ is null.

2. Consider now the case of an edge $e$ of $C$ such that $\e^K_C(e)=-1$. This time, $\tilde f(K)$ along $\tilde\gamma_e$ defines a simple
close curve around the origin (see Figure \ref{3cases}). Its contribution to $\tilde\omega$ is equal to $1\pmod{2}$.

3. Let $v$ be a vertex of $C$ with a dimer of $D$ sticking out of $v$ to the left of $C$. Then, the vector field $\tilde f(K)$ along
$\tilde\gamma_v$ induces a simple closed curve around the origin, so its contribution to $\tilde\omega$ is equal to $1\pmod{2}$.
The case illutrated in Figure \ref{3cases} is when $K$ orients the dimer from $v$
to its other boundary vertex. The other case is similar.

Gathering all the pieces, the winding number of $\tilde f(K)$ along $\tilde\gamma$ is equal to $\prod_{e\in C}\e^K_C(e)(-1)^{\ell_D(C)}$.
This concludes the proof of Theorem \ref{thm:spin}.\qed

\medskip

Using Johnson's theorem, we have the following immediate consequence of Corollary \ref{cor:quadratic}.

\begin{cor}\label{cor:spin}
Any dimer configuration $D$ on a surface graph $\GSS$ induces an isomorphism of affine $H^1(\Sigma;\Z_2)$-spaces
\[
\psi_D\colon(\K(\Gamma)/\sim)\longrightarrow {\mathcal{S}}(\SI)
\]
from the set of equivalence classes of Kasteleyn orientations on $\GSS$ onto the set of spin structures on $\SI$.
Furthermore, $\psi_D=\psi_{D'}$ if and only if $D$ and $D'$ are equivalent dimer configurations.\qed
\end{cor}

\section{Pffafian formulae for the partition function and correlation functions}\label{section:Pfaff}

\subsection{The Kasteleyn matrix and its Pfaffian}

Let $K$ be a Kasteleyn orientation on a surface graph $\Gamma\subset\Sigma$ with an even number of vertices. Enumerate these vertices
by $1,2,\dots, N=2n$. The {\em Kasteleyn matrix\/} is the $2n\times 2n$ skew-symmetric matrix $A^K(\Gamma;w)=A^K$ whose entry $a^K_{ij}$ is the
total weight
of all edges from $i$ to $j$ minus the total weight of all edges from $j$ to $i$. More formally,
\[
a^K_{ij}=\sum_{e}\e_{ij}^K(e)w(e),
\]
where the sum is on all edges $e$ in $\Gamma$ between the vertices $i$ and $j$, $\e_{ij}^K(e)=0$ if $i=j$, and
\[
\e^K_{ij}(e)=
\begin{cases}
\phantom{-}1 & \text{if $e$ is oriented by $K$ from $i$ to $j$;} \\
-1 & \text{otherwise,}
\end{cases}
\]
if $i\neq j$.

Consider a dimer configuration $D$ on $\Gamma$ given by edges $e_1,\dots,e_n$ matching vertices $i_\ell$ and $j_\ell$ for $\ell=1,\dots,n$.
It determines an equivalence class
of permutations $\sigma\colon (1,\dots, 2n)\mapsto (i_1,j_1,\dots,i_n,j_n)$ with respect to permutations of pairs $(i_\ell,j_\ell)$ and transpositions
$(i_\ell,j_\ell)\mapsto (j_\ell,i_\ell)$. We will write this as $\sigma\in D$. Given such a permutation $\sigma$, define
\[
\e^K(D)=(-1)^\sigma\prod_{\ell=1}^n\e^K_{i_\ell j_\ell}(e_\ell),
\]
where $(-1)^\sigma$ denotes the sign of the permutation $\sigma$. Note that this expression does not depend on the choice of $\sigma\in D$,
but only on the dimer configuration $D$.

\begin{thm}\label{quadratic}
Let $\Gamma\subset\Sigma$ be a surface graph with an even number of vertices.
For any dimer configuration $D_0$ on $\Gamma$ and any Kasteleyn orientation $K$ on $\Gamma\subset\Sigma$,
\begin{equation}\label{linear}
\e^K(D_0)\Pf(A^K)=\sum_{\alpha\in H_1(\Sigma;\Z_2)}(-1)^{q^K_{D_0}(\alpha)}\,Z_\alpha(D_0),
\end{equation}
where $q^K_{D_0}$ is the quadratic form on $H_1(\Sigma;\Z_2)$ associated to $K$ and $D_0$,
and
\[
Z_\alpha(D_0)=\sum_D w(D),
\]
the sum being on all dimer configurations $D$ such that $\Delta(D_0,D)=\alpha$.
\end{thm}
\begin{proof}
Recall that the Pfaffian of a skew-symmetric matrix $A=(a_{ij})$ of size $2n$ is given by
\[
\Pf(A)=\sum_{[\sigma]\in\Pi}(-1)^\sigma a_{\sigma(1)\sigma(2)}\cdots a_{\sigma(2n-1)\sigma(2n)},
\]
where the sum is on the set $\Pi$ of matchings of $\{1,\dots,2n\}$. Therefore,
\begin{align*}
\Pf(A^K)&=\sum_{[\sigma]\in\Pi}(-1)^\sigma a^K_{\sigma(1)\sigma(2)}\cdots a^K_{\sigma(2n-1)\sigma(2n)}\\
&=\sum_{[\sigma]\in\Pi}(-1)^\sigma\sum_{e_1}\e^K_{\sigma(1)\sigma(2)}(e_1)w(e_1)\cdots\sum_{e_n}\e^K_{\sigma(2n-1)\sigma(2n)}(e_n)w(e_n)\\
&=\sum_D (-1)^\sigma\prod_{\ell=1}^n\e^K_{\sigma(2\ell-1)\sigma(2\ell)}(e_\ell)w(e_\ell)\\
&=\sum_D \e^K(D)w(D),
\end{align*}
where the sum is on all dimer configurations $D$ on $\Gamma$. Hence,
\[
\e^K(D_0)\Pf(A^K)=\sum_D \e^K(D_0)\e^K(D)w(D).
\]
Let us denote by $e_1,\dots,e_n$ the edges of $\Gamma$ occupied by dimers of $D$, and by $e^0_1,\dots,e^0_n$ the edges
occupied by dimers of $D_0$. Fix permutations $\sigma$ and $\tau$ representing the dimer configurations $D$ and $D_0$, respectively,
and set $\nu=\tau\circ\sigma^{-1}$. By definition,
\begin{align*}
\e^K(D_0)\e^K(D)&=(-1)^{\tau}\prod_{\ell=1}^n\e^K_{\tau(2\ell-1)\tau(2\ell)}(e^0_\ell)
        \cdot(-1)^{\sigma}\prod_{\ell=1}^n\e^K_{\sigma(2\ell-1)\sigma(2\ell)}(e_\ell)\\
&=(-1)^\nu\prod_{\ell=1}^n\e^K_{\nu(\sigma(2\ell-1))\nu(\sigma(2\ell))}(e^0_\ell)\e^K_{\sigma(2\ell-1)\sigma(2\ell)}(e_\ell).
\end{align*}
Note that the permutation $\nu$ depends on the choice of $\sigma\in D$ and $\tau\in D_0$, but it always brings the perfect matching $D$ to $D_0$.
Moreover, one can choose representatives $\sigma\in D$ and $\tau\in D_0$ such that $\nu$ is the counter-clockwise rotation by one edge of every
$(D_0,D)$-composition cycle $C_1,\dots,C_m$. For this particular choice of representatives, we have
\[
\e^K(D_0)\e^K(D)=(-1)^{\sum_{i=1}^m({\mathrm{length}}(C_i)+1)}\prod_{i=1}^m\e^K(C_i)=\prod_{i=1}^m(-\e^K(C_i)).
\]
Here, we use the fact that the length of a permutation cycle is the length of the corresponding composition cycle,
and that the length of each composition cycle is even. Recall the quadratic form $q^K_{D_0}$ of Theorem \ref{thm:spin}.
Since the $C_i$'s are disjoint $(D_0,D)$-composition cycles, $C_i\cdot C_j=0$ and $\ell_{D_0}(C_i)=0$ for all $i,j$. Therefore,
\[
\prod_{i=1}^m(-\e^K(C_i))=(-1)^{q^K_{D_0}(\Delta(D_0,D))}.
\]
The theorem follows.
\end{proof}

\subsection{The partition function}

Given a dimer configuration $D_0$, Theorem \ref{quadratic}
provides $2^{2g}$ linear equations (one for each equivalence class
of Kasteleyn orientation) with $2^{2g}$ unknowns (the functions
$Z_\alpha(D_0)$). We want to use these equations to express the
dimer partition function $Z=\sum_\alpha Z_\alpha(D_0)$ in terms of
Pfaffians.

Let $V\times V\to\Z_2$, $(\alpha,\beta)\mapsto\alpha\cdot\beta$ be
a non-degenerate bilinear form on a $\Z_2$-vector space $V$.
Recall that the Arf invariant of a quadratic form $q\colon V\to\Z_2$
on $(V,\cdot)$ is given by
\[
\A(q)=\frac{1}{|V|}\sum_{\alpha\in V}(-1)^{q(\alpha)}.
\]

\begin{lemma}\label{lemma:Arf}
Let $q,q'$ be two quadratic forms on $(V,\cdot)$.
Then,
\[
\A(q)\A(q')=(-1)^{q(\Delta)}=(-1)^{q'(\Delta)},
\]
where $\Delta\in V$ satisfies $(q+q')(\alpha)=\Delta\cdot\alpha$
for all $\alpha\in V$.
\end{lemma}
\begin{proof}
First note that $q+q'$ is a linear form on $V$.
Since the bilinear form $\cdot$ is non-degenerate, there exists
$\Delta\in V$ such that $(q+q')(\alpha)=\alpha\cdot\Delta$ for
all $\alpha\in V$. Furthermore, $(q+q')(\Delta)=\Delta\cdot\Delta=0$, so
$(-1)^{q(\Delta)}=(-1)^{q'(\Delta)}$. Let us now compute the
product of the Arf invariants:
\[
\A(q)\A(q')
=\frac{1}{|V|}\sum_{\alpha,\beta\in V}(-1)^{q(\alpha)+q'(\beta)}
=\frac{1}{|V|}\sum_{\alpha,\beta\in V}(-1)^{q(\alpha)+q'(\beta+\Delta)}.
\]
Using the equality
\begin{align*}
q(\alpha)+q'(\beta+\Delta)&=q(\alpha)+q'(\beta)+q'(\Delta)+\beta\cdot\Delta\\
&=q(\alpha)+q(\beta)+q'(\Delta)=q(\alpha+\beta)+\alpha\cdot\beta+q(\Delta),
\end{align*}
we obtain
\begin{align*}
\A(q)\A(q')&=\frac{(-1)^{q(\Delta)}}{|V|}\sum_{\alpha,\beta}(-1)^{q(\alpha+\beta)+\alpha\cdot\beta}\\
&=\frac{(-1)^{q(\Delta)}}{|V|}\Big(\sum_\alpha 1+ \sum_{\alpha\neq \beta}(-1)^{q(\alpha+\beta)+\alpha\cdot\beta}\Big).
\end{align*}
We are left with the proof that the latter sum is zero:
\[
\sum_{\alpha\neq \beta}(-1)^{q(\alpha+\beta)+\alpha\cdot\beta}=
\sum_{\gamma\neq 0}(-1)^{q(\gamma)}\sum_{\alpha+\beta=\gamma}(-1)^{\alpha\cdot\beta}=
\sum_{\gamma\neq 0}(-1)^{q(\gamma)}(n^0_\gamma-n^1_{\gamma}),
\]
where $n_\gamma^i$ is the cardinality of the set
\[
N^i_\gamma=\{(\alpha,\beta)\in V\times V\,|\,\mbox{$\alpha+\beta=\gamma$ and $\alpha\cdot\beta=i$}\}
\]
for $i=0,1$. Since $\gamma\neq 0$ and $(V,\cdot)$ is
non-degenerate, there exists $x\in V$ such that $x\cdot\gamma=1$.
Then, the map $(\alpha,\beta)\mapsto (\alpha+x,\beta+x)$ induces
a bijection $N^0_\gamma\to N^1_\gamma$. Hence
$n^0_\gamma=n^1_\gamma$ for all $\gamma\neq 0$, and the lemma is
proved.
\end{proof}

Using this lemma, it is easy to check that if $\dim(V)=2n$,
then there are exactly $2^{2n-1}+2^{n-1}$ quadratic forms on $(V,\cdot)$
with Arf invariant $1$ and $2^{2n-1}-2^{n-1}$ forms with Arf invariant $-1$.

\begin{thm}\label{thm:partition}
The partition function of a dimer model on
a surface graph $\Gamma\subset\Sigma$ for a closed surface of
genus $g$ is given by the formula
\begin{equation}\label{Z}
Z=\frac{1}{2^{g}}\sum_{[K]}\A(q^{K}_{D_0})\e^{K}(D_0)\Pf(A^{K}),
\end{equation}
where the sum is taken over all equivalence classes of Kasteleyn
orientations. Each summand is defined for a Kasteleyn orientation but it
depends only on its equivalence class. Furthermore, the sign
$\A(q^K_{D_0})\e^{K}(D_0)$ does not depend on $D_0$.
\end{thm}

\begin{proof}
Recall that there is a free, transitive action of $H^1(\Sigma;\Z_2)$ on the
set of equivalence classes of Kasteleyn orientations on
$\Gamma\subset\Sigma$ (Theorem \ref{thm:aff}). Let us denote by
$K_\phi$ the result of the action of $\phi\in H^1(\Sigma;\Z_2)$ on
a Kasteleyn orientation $K$. Then, by the first part of Proposition \ref{prop:comp},
\[
(-1)^{(q^K_{D_0}+q^{K_\phi}_{D_0})(\alpha)}=(-1)^{\phi(\alpha)}=:\chi_\alpha(\phi).
\]
Here $\chi_\alpha$'s are characters of  irreducible
representations of $H^1(\Sigma;\Z_2)$. 

By equation (\ref{linear}),
\[
\e^{K_\phi}(D_0)\Pf(A^{K_\phi})=\sum_{\alpha\in H_1(\Sigma;\Z)}(-1)^{q^{K_\phi}_{D_0})(\alpha)}Z_\alpha(D_0)
\]
for all $\phi\in H^1(\Sigma;\Z)$. Multiplying these equations by $(-1)^{q^{K_\phi}_{D_0}(\beta)}$ and taking the sum
over all $\phi\in H^1(\Sigma;\Z)$, we get
\begin{align*}
\sum_\phi(-1)^{q^{K_\phi}_{D_0}(\beta)}\e^{K_\phi}(D_0)\Pf(A^{K_\phi})
&=\sum_\phi\sum_\alpha (-1)^{q^{K_\phi}_{D_0}(\alpha)+q^{K_\phi}_{D_0}(\beta)}Z_\alpha(D_0)\\
&=\sum_\alpha(-1)^{q^K_{D_0}(\alpha)+q^K_{D_0}(\beta)}\sum_\phi\chi_\alpha(\phi)\chi_\beta(\phi)Z_\alpha(D_0).
\end{align*}
Using the orthogonality formula $\sum_\phi\chi_\alpha(\phi)\chi_\beta(\phi)=2^{2g}\delta_{\alpha\beta}$, we obtain
\[
Z_\alpha(D_0)=\frac{1}{2^{2g}}\sum_\phi(-1)^{q^{K_\phi}_{D_0}(\alpha)}\e^{K_\phi}(D_0)\Pf(A^{K_\phi}).
\]
Therefore, the partition function $Z=\sum_\alpha Z_\alpha(D_0)$ is given by
\[
Z=\frac{1}{2^{g}}\sum_{\phi\in H^1(\Sigma;\Z_2)}\sigma^{K_\phi}\Pf(A^{K_\phi}),
\]
where
\[
\sigma^{K_\phi}=\e^{K_\phi}(D_0)\frac{1}{2^g}\sum_{\alpha\in
H_1(\Sigma;\Z_2)}(-1)^{q^{K_\phi}_{D_0}(\alpha)}=\e^{K_\phi}(D_0)\A(q^{K_\phi}_{D_0}).
\]
This gives the formula
\[
Z=\frac{1}{2^{g}}\sum_{\phi\in H^1(\Sigma;\Z_2)}\A(q^{K_\phi}_{D_0})\e^{K_\phi}(D_0)\Pf(A^{K_\phi}).
\]
Since $H^1(\Sigma,\Z_2)$ acts transitively and freely on the
equivalence classes of Kasteleyn orientations, equality $(\ref{Z})$ follows.

If $K$ and $K'$ are equivalent Kasteleyn orientations, then $q^K_{D_0}=q^{K'}_{D_0}$ so $\A(q^K_{D_0})=\A(q^{K'}_{D_0})$.
On the other hand, $\e^K(D_0)=(-1)^\mu\e^{K'}(D_0)$ and $\Pf(A^K)=(-1)^\mu\Pf(A_{K'})$, where $\mu$ is the number of vertices of
$\Gamma$ around which the orientation was flipped. Therefore, the summand $\A(q^{K}_{D_0})\e^{K}(D_0)\Pf(A^{K})$ does not
depend on the choice of the representative in the equivalence class $[K]$.

Let us finally check that the sign $\A(q^{K}_{D_0})\e^{K}(D_0)$ does not depend on $D_0$. Let $D$ be another dimer
configuration on $\Gamma$. By Proposition \ref{prop:comp}, Lemma \ref{lemma:Arf} and the proof of Theorem \ref{quadratic},
\[
A(q^{K}_{D_0})\e^{K}(D_0)A(q^{K}_D)\e^{K}(D)=(-1)^{q^K_D(\Delta(D_0,D))}\A(q^{K}_{D_0})A(q^{K}_D)=1.
\]
This concludes the proof of the theorem.
\end{proof}

\subsection{Local correlation functions}

In order to express local correlation functions
$<\sigma_{e_1}\cdots\sigma_{e_k}>$ as combinations of Pfaffians,
let us recall several facts of linear algebra.

Let $A=(a_{ij})$ be a matrix of size $2n$. Given an ordered subset
$I$ of the ordered set $\alpha=(1,\dots,2n)$, let $A_I$ denote the
matrix obtained from $A$ by removing the $i^\mathrm{th}$ row and
the $i^\mathrm{th}$ column for all $i\in I$. Also, let
$(-1)^{\sigma(I)}$ denote the signature of the permutation which
sends $\alpha$ to the ordered set $I(\alpha\backslash I)$. If $A$
is skew-symmetric, then for all ordered set of indices
$I=(i_1,j_1,\dots,i_k,j_k)$,
\[
\frac{\partial^k\Pf(A)}{\partial a_{i_1j_1}\cdots\partial
a_{i_kj_k}}=(-1)^{\sigma(I)}\Pf(A_I).
\]
Furthermore, if $A$ is invertible, then $\Pf(A)\neq 0$ and
\[
(-1)^{\sigma(I)}\Pf(A_I)=(-1)^k\Pf(A)\Pf((A^{-1})_{\alpha\backslash
I}).
\]

So, let $e_1,\dots,e_k$ be edges of the graph $\G$, and let
$i_\ell,j_\ell$ be the two boundary vertices of $e_\ell$ for
$\ell=1,\dots,k$. For simplicity, we shall assume that $\Gamma$ has
no multiple edges. Finally, set $I=(i_1,j_1,\dots,i_k,j_k)$.
Applying the identities above to the Kasteleyn matrices
$A^\phi=A^{K_\phi}$, Theorem \ref{thm:partition} gives
\begin{align*}
\prod_{\ell=1}^k w(e_\ell)\frac{\partial^k Z}{\partial w(e_1)\cdots\partial w(e_k)}&=
\frac{1}{2^g}\sum_\phi\sigma^{K_\phi}\prod_{\ell=1}^k a^\phi_{i_\ell
j_\ell}\frac{\partial^k}{\partial a^\phi_{i_1j_1}\cdots\partial
a^\phi_{i_kj_k}}\Pf(A^\phi)\\
&=\frac{1}{2^g}\sum_\phi\sigma^{K_\phi}\prod_{\ell=1}^k a^\phi_{i_\ell
j_\ell}(-1)^{\sigma(I)}\Pf(A^\phi_I)
\end{align*}

If the Kasteleyn matrix is invertible for any Kasteleyn orientation of
$\Gamma$, this expression is equal to
\[
\frac{(-1)^k}{2^g}\sum_\phi\sigma^{K_\phi}\prod_{\ell=1}^k a^\phi_{i_\ell
j_\ell}\Pf(A^\phi)\Pf((A^\phi)^{-1}_{\alpha\backslash I}).
\]
Since
\[<\sigma_{e_1}\cdots\sigma_{e_k}>=\frac{w(e_1)\cdots
w(e_k)}{Z}\frac{\partial^kZ}{\partial w(e_1)\cdots\partial w(e_k)},
\]
the correlation functions are given by
\[
<\sigma_{e_1}\cdots\sigma_{e_k}>=
(-1)^k\frac{\sum_\phi\sigma^{K_\phi}\Pf(A^\phi)\prod_\ell
a^\phi_{i_\ell j_\ell}\Pf((A^\phi)^{-1}_{\alpha\backslash I})}
        {\sum_\phi\sigma^{K_\phi}\Pf(A^\phi)}.
\]
if the Kasteleyn matrix is invertible for all possible Kasteleyn orientations of $\Gamma$.

\medskip

If $\G$ is a planar graph, then the Kasteleyn matrix is always invertible since its Pfaffian
is equal to the partition function. Therefore,
\[
<\sigma_{e_1}\cdots\sigma_{e_k}>=(-1)^k a^K_{i_1j_1}\cdots
a^K_{i_kj_k}\Pf((A^K)^{-1}_{\alpha\backslash I}),
\]
where $K$ is any Kasteleyn orientation on $\G\subset S^2$.

On the other hand, there are graphs where some Kasteleyn matrix is not invertible.
For example, consider a square lattice on a torus. Then, the Kasteleyn
matrix corresponding to the spin structure with Arf invariant $-1$ is not invertible (see \cite{McCoyW}).

\appendix

\section{Dimers and Grassman integrals}

\subsection{Grassman integrals and Pfaffians}

Let $V$ be an $n$-dimensional vector space. Its exterior algebra $\wedge
V=\oplus_{k=0}^n\wedge^k V$ is called the {\em Grassman algebra of $V$\/}.
The choice of a linear basis in $V$ induces an isomorphism between $\wedge V$
and the algebra generated by elements $\phi_1,\dots, \phi_n$ with defining relations
$\phi_i\phi_j=-\phi_j\phi_i$. The isomorphism identifies the linear basis in $V$
with the generators $\phi_i$.

Choose an orientation on $V$. Together with the basis in $V$, this defines a basis in the top exterior
power of $V$. The {\em integral over the Grassman algebra of $V$\/} of an element $a\in\wedge V$
is the coordinate of $a$ in the top exterior
power of $V$ with respect to this basis. It is denoted by
\[
\int a\,d\phi.
\]

In physics, elements $\phi$ are called {\em Fermionic fields\/}. More
precisely, they are called neutral fermionic fields (or neutral fermions).
They are called charged fermions if there is an action of $U(1)$ on $V$.

Recall that the Pfaffian of a skew symmetric matrix $A$ of even size $n$ is given by
\[
\Pf(A)=\frac{1}{2^{\frac{n}{2}} \left ({\frac{n}{2}}\right )!}
\sum_{\sigma\in S_n}(-1)^\sigma a_{\sigma(1)\sigma(2)}\cdots a_{\sigma(n-1)\sigma(n)},
\]
where the sum is over all permutations $\sigma\in S_n$. (This formula is easily seen to be equivalent to the one
stated in Section \ref{section:Pfaff}.) Expanding the exponent into a power series, one gets the following identity
in $\wedge V$:
\[
\pi\Big(\exp\Big(\frac{1}{2}\sum_{i,j=1}^na_{ij}\phi_i\wedge \phi_j\Big)\Big)=\Pf(A)\,\phi_1\wedge\dots\wedge\phi_n,
\]
where $\pi\colon\wedge V\to \wedge^nV$ is the projection to the top exterior power.
In terms of Grassmann integral, this can be expressed as
\begin{equation}\label{majgauss}
\int \exp\Big(\frac{1}{2}\sum_{i,j=1}^n\phi_i a_{ij}\phi_j\Big)d\phi=\Pf(A).
\end{equation}

\bigskip

Let us now assume that the space $V$ is polarized, i.e. that $V=W\oplus W^*$ with $W$ some vector
space and $W^*$ its dual. Then, the Grassman algebra of $V$
is isomorphic to the tensor product of Grassman algebras for $W$
and for $W^*$ in the category of super-vector spaces. That is, if
the dimension of $W$ is $k$, the Grassman algebra of $V$ is
isomorphic to the algebra generated by $\psi_i, \psi_i^*,
i=1,\dots k$ with defining relations $\psi_i\psi_j=-\psi_j\psi_i$,
$\psi_i\psi_j^*=-\psi_j^*\psi_i$, and $\psi_i^*\psi_j^*=-\psi_j^*\psi_i^*$.
Such an isomorphism is specified by the choice of a linear basis in $W$. Note that such a choice
induces a basis in $V$ (take the dual basis in $W^*$) and an orientation on $V$ given by the
ordering $\psi_1,\dots,\psi_n,\psi^*_1,\dots,\psi_n^*$.
The Grassmann integral of an element $a\in\wedge V$ with respect to this choice of basis in $W$ is denoted by
\[
\int a\,d\psi d\psi^*.
\]
Expanding the exponent, we obtain
\begin{equation}\label{det}
\int \exp\Big(\sum_{i,j=1}^k \psi_i a_{ij}\psi^*_j\Big)d\psi d\psi^*=(-1)^{\frac{k(k-1)}{2}}\Det(A).
\end{equation}
Comparing this formula with (\ref{majgauss}), we obtain the following well known identity
\[
\Pf\left(\begin{array}{cc}0&A\\-A^t&0\end{array}\right)=(-1)^{\frac{k(k-1)}{2}}\Det(A).
\]

\bigskip

Now consider the space $U=V\oplus V^*$. Let $\phi_1,\dots,\phi_n$ be a basis in $V$ and
$\phi^*_1,\dots,\phi^*_n$ be the dual basis in $V^*$. Using the change of variables
$\chi_i=\frac{\phi_i+\sqrt{-1}\phi^*_i}{\sqrt{2}}$, $\chi^*_i=\frac{\phi_i-\sqrt{-1}\phi^*_i}{\sqrt{2}}$,
together with the equalities (\ref{majgauss}) and (\ref{det}), we obtain 
\begin{align*}
\Pf(A)^2
&=\int\exp\Big(\frac{1}{2}\sum_{i,j=1}^na_{ij}\phi_i\phi_j+\frac{1}{2}\sum_{i,j=1}^na_{ij}\phi^*_i\phi^*_j\Big)d\phi d\phi^*\\
&=(-1)^{\frac{n(n-1)}{2}}\int\exp\Big(\sum_{i,j=1}^n \chi_i a_{ij}\chi^*_j\Big)d\chi d\chi^*\\
&=\Det(A).
\end{align*}

One can easily derive other Pfaffian identites in a similar way .

\subsection{Dimer models on graphs and Grassman integrals}

Let $\GSS$ be a surface graph, and let $A^K=(a^K_{ij})$ be the Kasteleyn matrix associated with
a Kasteleyn orientation $K$ on $\GSS$.
By Theorem \ref{thm:partition} and identity (\ref{majgauss}), the partition function for dimers on $\G$ is given by
\[
Z=\frac{1}{2^{g}}\sum_{[K]}\A(q^{K}_{D_0})\e^{K}(D_0)\int\exp\Big(\frac{1}{2}\sum_{i,j\in V(\G)}\phi_i a^K_{ij} \phi_j\Big)d\phi,
\]
where $V(\G)$ denotes the set of vertices of $\G$.

Recall that $\e^K(D_0)=(-1)^\sigma\prod_{\ell=1}^n\e^K_{i_\ell j_\ell}$,
where the dimer configuration $D_0$ matches vertices $i_\ell$ and $j_\ell$ for $\ell=1,\dots,n$,
and $\sigma$ is the permutation $(1,\dots, 2n)\mapsto (i_1,j_1,\dots,i_n,j_n)$. Taking into account the identity 
\[
(-1)^\sigma\prod_{\ell=1}^n\e^K_{i_\ell j_\ell}\,d\phi_1\dots d\phi_{2n}=\prod_{\ell=1}^n\e^K_{i_\ell j_\ell}\,d\phi_{i_\ell}d\phi_{j_\ell},
\]
the formula for the partition function can be written as
\[
Z=\frac{1}{2^{g}}\sum_{[K]}\int\exp\Big(\frac{1}{2}\sum_{i,j\in V(\G)}\phi_i a^K_{ij} \phi_j\Big)D^K\phi,
\]
where 
\[
D^K\phi=\A(q^{K}_{D_0})\e^{K}(D_0)\,d\phi=\A(q^{K}_{D_0})
\prod_{\ell=1}^n\e^K_{i_\ell j_\ell}\,d\phi_{i_\ell}d\phi_{j_\ell}.
\]

Similarly, local correlation functions can be written as
\[
<\sigma_{e_1}\cdots\sigma_{e_k}>=\frac{1}{2^gZ}\sum_{[K]}\int \exp\Big(\frac{1}{2}\sum_{i,j\notin I}\phi_i a^K_{ij}\phi_j\Big)
\prod_{l=1}^k a^K_{i_lj_l}\phi_{i_l}\phi_{j_l}D^K\phi .
\]
Here $i_l$ and $j_l$ are the boundary vertices of the edge $e_l$, and $I=(i_1,j_1,\dots,i_k,j_k)$.

\bibliographystyle{amsplain}

\begin{thebibliography}{10}

\bibitem{AGBMNV}
L. Alvarez-Gaumé, J.-B. Bost, G. Moore, P. Nelson and C. Vafa,
{\em Bosonization on higher genus Riemann surfaces\/},
Comm. Math. Phys. \textbf{112} (1987), no. 3, 503--552.

\bibitem{CSMCoy}
R. Costa-Santos and B. McCoy, {\em Dimers and the critical Ising model on lattices of genus $>1$\/},
Nuclear Phys. B  \textbf{623} (2002),  no. 3, 439--473.

\bibitem{CKP}
H. Cohn, R. Kenyon and J. Propp, {\em  A variational principle for domino tilings\/},
J. Amer. Math. Soc. \textbf{14} (2001),  no. 2, 297--346.

\bibitem{Mis}
N. Dolbilin,  Yu. Zinovyev, A. Mishchenko, M. Shtanko and M.  Shtogrin,
{\em Homological properties of two-dimensional coverings of lattices on surfaces\/},
(Russian)  Funktsional. Anal. i Prilozhen. \textbf{30} (1996),  no. 3, 19--33, 95;
translation in  Funct. Anal. Appl. \textbf{30} (1996), no. 3, 163--173 (1997).

\bibitem{Joh}
D. Johnson, {\em Spin structures and quadratic forms on surfaces\/}, J. London Math. Soc. (2) \textbf{22} (1980), no. 2,
365--373.

\bibitem{Kast1}
W. Kasteleyn, {\em Dimer statistics and phase transitions\/},
J. Mathematical Phys. \textbf{4} (1963) 287--293.

\bibitem{Kast2}
W. Kasteleyn, Graph theory and crystal physics. 1967
Graph Theory and Theoretical Physics  pp. 43--110 Academic Press, London.

\bibitem{Ken}
R. Kenyon, {\em The Laplacian and Dirac operators on critical planar graphs\/},
Invent. Math. \textbf{150} (2002), no. 2, 409--439.

\bibitem{KenOk}
R. Kenyon and A. Okounkov, {\em Planar dimers and Harnack curves\/},
Duke Math. J. \textbf{131} (2006), no. 3, 499--524.

\bibitem{KenOkS}
R. Kenyon, A. Okounkov and S. Sheffield, {\em Dimers and amoebae\/},
Ann. of Math. (2) \textbf{163} (2006), no. 3, 1019--1056.

\bibitem{Kup}
G. Kuperberg, {\em An exploration of the permanent-determinant method\/},
Electron. J. Combin.  \textbf{5}  (1998), Research Paper 46, 34 pp. (electronic).

\bibitem{L}
A. Galluccio and M. Loebl, {\em On the theory of Pfaffian orientations.
I. Perfect matchings and permanents.}
Electron. J. Combin. \textbf{6} (1999), Research Paper 6, 18 pp. (electronic).

\bibitem{LP}
L. Lovasz and  M.D. Phummer, {\em Matching theory\/},
North-Holland Mathematics Studies, 121. Annals of Discrete Mathematics, 29.
North-Holland Publishing Co., Amsterdam.

\bibitem{M}
C. Mercat, {\em Discrete Riemann surfaces and the Ising model\/},
Comm. Math. Phys. \textbf{218} (2001), no. 1, 177--216.

\bibitem{McCoyW}
B. McCoy and T.T.  Wu, {\em The two-dimensional Ising model\/},
Harvard University Press, Cambridge Massachusetts, 1973.

\bibitem{T}
G. Tesler, {\em Matchings in graphs on non-orientable surfaces\/},
J. Combin. Theory Ser. B  \textbf{78} (2000),  no. 2, 198--231.
\end{thebibliography}

\end{document}